\documentclass[aps,prl,twocolumn,amsmath,amssymb,floatfix,longbibliography,superscriptaddress]{revtex4-2}

\usepackage[T1]{fontenc}
\usepackage{blindtext}
\usepackage{lineno}
\usepackage{graphicx}
\usepackage{dcolumn}
\usepackage{bm}
\usepackage[normalem]{ulem} 
\usepackage{amsmath,bm}
\usepackage{amsfonts}
\usepackage{amssymb,enumerate}
\usepackage{graphicx,hyperref}
\usepackage{amssymb}
\usepackage{epstopdf,booktabs}
\usepackage{empheq}
\usepackage{fancybox}
\usepackage{ulem}
\usepackage{amssymb}
\usepackage{upgreek}
\usepackage[usenames,dvipsnames]{color}
\usepackage[utf8]{inputenc}
\usepackage{natbib}
\usepackage{soul}
\hypersetup{hidelinks,colorlinks=true,citecolor=blue,linkcolor=blue,urlcolor=black}

\newcommand*{\rom}[1]{\expandafter\@slowromancap\romannumeral #1@}
\begin{document}

\title{Flying focus with arbitrary directionality for spatiotemporal control of laser pulses}

\author{Sida Cao}
\email{sidacao@stanford.edu}
\affiliation{Department of Mechanical Engineering, Stanford University, Stanford, California 94305, USA}

\author{Devdigvijay Singh}
\affiliation{Department of Mechanical Engineering, Stanford University, Stanford, California 94305, USA}

\author{Lavonne S. Mack}
\affiliation{Laboratory for Laser Energetics, University of Rochester, Rochester, New York 14623, USA}

\author{John P. Palastro}
\affiliation{Laboratory for Laser Energetics, University of Rochester, Rochester, New York 14623, USA}

\author{Matthew R. Edwards}
\email{mredwards@stanford.edu}
\affiliation{Department of Mechanical Engineering, Stanford University, Stanford, California 94305, USA}


\date{\today}

\begin{abstract}

Flying focus techniques produce laser pulses whose focal points travel at arbitrary, controllable velocities. 
While this flexibility can enhance a broad range of laser-based applications, existing techniques constrain the motion of the focal point to the propagation direction of the pulse. 
Here, we introduce a flying focus configuration that decouples the motion of the focus from the propagation direction. 
A chirped laser pulse focused and diffracted by a diffractive lens and grating creates a focal point that can move both along and transverse to the propagation direction. 
The focal length of the lens, grating period, and chirp can be tuned to control the direction and velocity of the focus. 
Simulations demonstrate this control for a holographic configuration suited to high-power pulses, in which two off-axis pump beams with different focal lengths encode the equivalent phase of a chromatic lens and grating in a gas or plasma. 
For low-power pulses, conventional solid-state or adaptive optics can be used instead. 
Multi-dimensional control over the focal trajectory enables new configurations for applications, including laser wakefield acceleration of ions, nonlinear Thomson scattering, and surface-plasmon emission of THz radiation.

\end{abstract}

\maketitle

\section{Introduction}
Optical techniques for spatiotemporal control reshape the amplitude, phase, or polarization of a laser pulse by introducing correlations between its spatial and temporal degrees of freedom~\cite{hall2021free,yessenov2022space,zhan2024spatiotemporal,piccardo2025trends}. These techniques have enabled the generation of propagation-invariant wave packets with tunable group velocity and acceleration~\cite{kondakci2019optical,li2020velocity,yessenov2020free,li2020optical,yessenov2020accelerating,hall2022arbitrarily,li2022investigating,su2024temporally,su2024temporally,su2025space}, dynamically steered laser beams~\cite{shaltout2019spatiotemporal}, novel optical topologies~\cite{piccardo2022vortex, piccardo2023broadband,franco2023curve, shen2025free}, simultaneous spatial and temporal focusing~\cite{jing2016characteristics}, and spatiotemporal optical vortices~\cite{feng2023terawatt,hancock2024spatiotemporal,le2024self,chen2025generation}. 
Beyond providing fundamental insights into the allowable structures of light, spatiotemporal control holds promise for several applications, including biomedical imaging~\cite{vettenburg2014light,piksarv2017integrated,wang2017simultaneous}, pulse compression~\cite{bor1985group}, terahertz generation~\cite{hebling2002velocity,nugraha2019demonstration,wang2020tilted}, polarization control~\cite{cao2022vectorial}, attosecond pulse generation~\cite{vincenti2012attosecond}, nonlinear harmonic generation~\cite{liang2019enhancement}, attsecond electron sheet generation~\cite{sun2024generation}, $\gamma$-ray generation~\cite{sun2025isolated}, and temporally resolved single-shot measurements~\cite{grace2025single, kim2021single}. 

A subset of techniques for spatiotemporal control, known as “flying focus” methods, modify the focal time and location of each frequency, temporal slice, or annulus of a pulse to create an intensity peak that moves independently of the group velocity along the direction of propagation~\cite{sainte2017controlling,froula2018spatiotemporal,franke2019measurement,jolly2020controlling,caizergues2020phase,palastro2020dephasingless,ambat2023programmable,simpson2020nonlinear,simpson2022spatiotemporal,pigeon2023ultrabroadband,liberman2024use,li2024spatiotemporal}. These modifications can be realized through a variety of optical configurations, including a chromatic lens and chirped laser pulse~\cite{sainte2017controlling,froula2018spatiotemporal,jolly2020controlling,li2024spatiotemporal}; an axilens and echelon or refractive doublet~\cite{caizergues2020phase, palastro2020dephasingless,pigeon2023ultrabroadband,liberman2024use}; a deformable mirror and spatial light modulator ~\cite{ambat2023programmable}; or nonlinear optical processes~\cite{simpson2020nonlinear,simpson2022spatiotemporal}.
The arbitrary and tunable velocity of the resulting intensity peak has provided a new approach to enhancing laser wakefield acceleration~\cite{palastro2020dephasingless,caizergues2020phase,geng2024efficient}, plasma-based parametric amplification~\cite{turnbull2018raman,wu2022self}, THz, extreme ultraviolet, and x-ray generation~\cite{howard2019photon,Ramsey2022,kabacinski2023spatio,Ye2023,simpson2024spatiotemporal,fu2025steering}, and signatures of strong-field quantum electrodynamics~\cite{formanek2022radiation,jin2023enhancement,formanek2024signatures}. 

Despite the flexibility afforded by flying-focus methods, existing implementations constrain the motion of the intensity peak to the propagation direction of the pulse. An arbitrary-velocity intensity peak capable of traveling in any direction would offer greater versatility, opening new geometries for laser-matter interactions. 
For instance, such peaks could be used to tune the undulator period while extending the interaction length in a laser-driven free-electron laser~\cite{ramsey2025x}, extend the interaction length and enhance x-ray emission from nonlinear Thomson scattering in a non-backscattering geometry for easier detection~\cite{Ramsey2022,ye2023enhanced}, excite surface plasmons for THz generation in non-normal geometries~\cite{welsh2007terahertz},  and enhance compact ion acceleration to GeV energies~\cite{gong2024laser}.

\begin{figure*}
    \centering
    \includegraphics[width=1\linewidth]{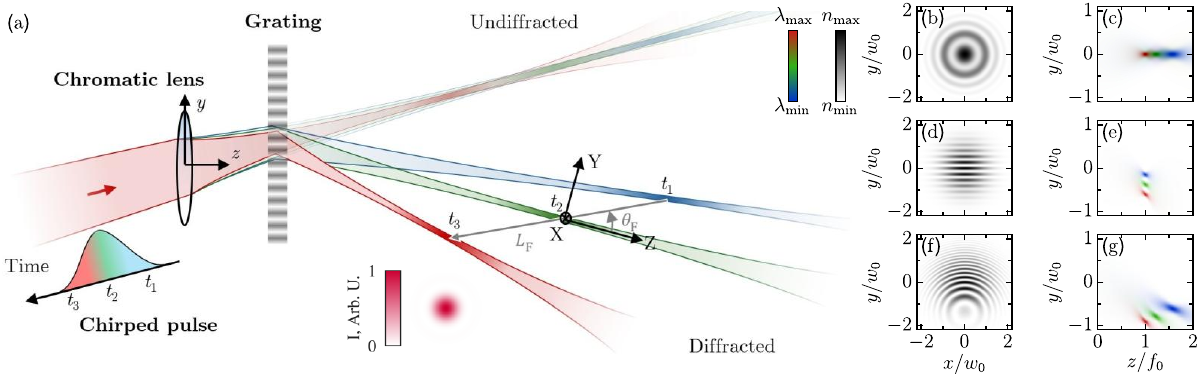}
    \caption{Schematic of a flying focus with arbitrary directionality. (a) A chirped pulse propagates through a chromatic lens and a grating, creating a flying focus at an angle $\theta_\mathrm{F}$ with respect to the pulse propagation direction with a focal range $L_\mathrm{F}$.
    (b) A diffractive lens disperses the frequencies longitudinally, producing (c) a focus that travels along the same direction as the pulse.
    (d) A diffraction grating disperses the frequencies transversely, producing (e) a focus that travels perpendicular to the propagation direction of the pulse.
    (f) A diffractive lens and grating combination disperses the frequencies both longitudinally and transversely, producing (g) a focus that travels at an angle $\theta_\mathrm{F}$ with respect to the propagation direction.
    The inset shows an example of the transverse intensity profile of the diffracted focal spot.}
    \label{fig:off-axis zone plate schematic}
\end{figure*}

In this work, we introduce a flying-focus method for producing intensity peaks that travel at any velocity in any direction. The optical configuration consists of a chirped laser pulse focused and diffracted by a diffractive lens and grating. The lens and grating determine the longitudinal and transverse focal location of each frequency, respectively, while the chirp determines the arrival time of each frequency at that location. The focal length of the lens, grating period, and chirp can be tuned to control the direction and velocity
of the resulting focus. Theoretical analysis and numerical simulations of this configuration demonstrate that the focus can travel at any angle with respect to the pulse propagation direction over distances far longer than a Rayleigh range. The simulations consider a holographic lens (zone plate) and grating, which can be generated in a gas or plasma suitable for high-power applications~\cite{michine2020ultra,edwards2022holographic,Edwards2022plasma,edwards2024structured, michel2024photochemically, piccardo2025trends, singh2025holographic, ou2026near, michel2026entropy}. However, the technique and analysis are general and can be applied to the standard solid-state or adaptive optics frequently used for low-power applications. 

\section{A Two-Dimensional Flying Focus}

Consider a laser pulse incident on a diffractive lens and transmission grating centered on the $z$-axis, as shown in Fig.~\ref{fig:off-axis zone plate schematic}(a). 
The transverse electric field of the pulse can be expressed as a superposition of its frequency components: 
\begin{equation}
E(\mathbf{x}, t) = \frac{1}{4\pi}\int \tilde{E}(\mathbf{x}, \omega)e^{i(kz - \omega t)}d\omega + \mathrm{c.c.},
\end{equation}
where $\tilde{E}(\mathbf{x},\omega) = A(\mathbf{x},\omega)e^{i\phi(\mathbf{x},\omega)}$ is the complex envelope of each component with amplitude $A(\mathbf{x},\omega) = |\tilde{E}(\mathbf{x},\omega)|$ and phase $\phi(\mathbf{x},\omega) = \mathrm{Arg}[\tilde{E}(\mathbf{x},\omega)]$, $k \equiv \omega/c$, and the integral is over positive frequencies. The vacuum wavelength associated with each frequency is $\lambda = 2 \pi c /\omega$. 

The pulse is incident at an angle $\theta_i$, vertically displaced a distance $y_i$ from the $z$-axis, 
has been focused by a preliminary achromatic lens, and has a spectral phase $\Phi(\omega)$. The incident phase of the pulse $\phi_i(\mathbf{x}_{\perp},\omega) \equiv \phi(\mathbf{x}_{\perp},z=0,\omega)$ is then
\begin{equation}
\phi_i(\mathbf{x}_{\perp},\omega)  = k\theta_i(y-y_i) - \frac{k[x^2+(y-y_i)^2]}{2f_i} + \Phi(\omega),
\end{equation}
where $z=0$ is the entrance plane of the diffractive lens and $f_i$ is the longitudinal distance from $z=0$ to the original focal point.

Upon entering the diffractive lens and thereafter, the envelope evolves according to the paraxial wave equation:
\begin{equation}
    \label{eq:paraxial wave equation}
    \left(2ik\partial_z + \nabla_\perp^2\right)\tilde{E}(\mathbf{x},\omega) = -k^2\left[n^2(\mathbf{x}) - 1\right]\tilde{E}(\mathbf{x},\omega),
\end{equation}
where $\nabla_\perp^2 = \partial_y^2 + \partial_x^2$ is the transverse Laplacian and $n(\mathbf{x})$ is the spatially dependent refractive index, which includes contributions from the lens and grating.
For thick optics, Eq.~\eqref{eq:paraxial wave equation} can be solved numerically to determine the envelope at the exit of the grating. For thin optics, as considered here, the envelope at the exit of the grating can be approximated as 
\begin{equation}
\label{eq:initialcond}
\tilde{E}(\mathbf{x}_{\perp},z=0^+,\omega) = A_i(\mathbf{x}_{\perp},\omega)e^{i\phi_i(\mathbf{x}_{\perp},\omega) + i\phi_\mathrm{C}(\mathbf{x}_{\perp},\omega)},
\end{equation}
where $A_i(\mathbf{x}_{\perp},\omega)\equiv A(\mathbf{x}_{\perp},z=0,\omega)$ is the incident amplitude, $z=0^+$ denotes a longitudinal location just after the lens and grating, and $\phi_\mathrm{C}(\mathbf{x}_{\perp},\omega) = \phi_\mathrm{L}(\mathbf{x}_{\perp},\omega) + \phi_\mathrm{G}(\mathbf{x}_{\perp},\omega)$ is the phase acquired by a frequency component upon traversing the lens (L) and grating (G).

The diffractive lens is designed to have a focal length $f_\mathrm{L0}$ at the central wavelength of the probe pulse $\lambda_0$. 
The phase applied to each frequency component is then
\begin{equation}
    \label{eq:quadratic phase}
    \phi_\mathrm{L}(\mathbf{x}_{\perp},\omega) = -\frac{k(x^2 + y^2)}{2f_\mathrm{L}(\lambda)},
\end{equation}
where $z = f_\mathrm{L}(\lambda) = f_\mathrm{L0}\lambda_0/\lambda$ is the focal location of the first-order diffraction for an initially collimated pulse (i.e., when $f_i \rightarrow \infty$). The transmission grating has a period $\Lambda$ in the $y$ direction, chosen to produce a first-order diffraction angle
\begin{equation}
    \label{eq:diffraction angle}
    \theta_\mathrm{D}(\lambda) = -\frac{\lambda}{\Lambda}.
\end{equation}
The phase applied by the grating is then
\begin{equation}
    \label{eq:phase diffraction grating}
    \phi_\mathrm{G}(\mathbf{x}_{\perp},\omega) = k\theta_\mathrm{D}(\lambda)y.
\end{equation}
The maximum diffraction efficiency occurs when the central wavelength of the pulse is incident at the Bragg angle $\theta_\mathrm{B} \approx \lambda_0/2\Lambda$.

Beyond the exit of the grating ($z>0^+$), $n(\mathbf{x}) = 1$, and the solution to Eq.~\eqref{eq:paraxial wave equation} is given by the Fresnel integral:
\begin{equation}
\label{eq:fresnel integral}
\begin{split}
    \tilde{E}(\mathbf{x},\omega) = \frac{k}{2\pi iz}\int \tilde{E}(\mathbf{x}_{\perp}',0^+,\omega)\exp{\left[i \tfrac{k}{2z}|\mathbf{x}_{\perp}-\mathbf{x}_{\perp}'|^2\right]
    }d\mathbf{x}_{\perp}',
\end{split}
\end{equation}
where $\tilde{E}(\mathbf{x}_{\perp}',0^+,\omega)$ is provided by Eq.~\eqref{eq:initialcond}. 
The focal location of each wavelength $\mathbf{x}_f(\lambda)$ can be determined by applying the method of stationary phase to the total phase in Eq.~\eqref{eq:fresnel integral}: $\phi_\mathrm{T} = \phi_i + \phi_\mathrm{C} + \tfrac{k}{2z}|\mathbf{x}_{\perp}-\mathbf{x}_{\perp}'|^2$.
At focus, all the rays that originate from the plane $z = 0^+$ are coherent with zero relative phase.
This condition can be formulated as:
\begin{subequations}
\label{eq:stationary phase}
    \begin{eqnarray}
    (\partial_{x'}\phi_\mathrm{T})|_{\mathbf{x} = \mathbf{x}_f} = 0,\\
    (\partial_{y'}\phi_\mathrm{T})|_{\mathbf{x} = \mathbf{x}_f} = 0,  
    \end{eqnarray}
\end{subequations}
for all $x'$ and $y'$. Substituting in for $\phi_\mathrm{T}$ and solving for $\mathbf{x}_{\perp}$ yields the first-order focal location for each wavelength:
\begin{subequations}
\label{eq:off-axis focal spot location and propagation direction}
\begin{eqnarray}
        z_f(\lambda)&=& \frac{\lambda_0f_if_\mathrm{L0}}{\lambda f_i + \lambda_0f_\mathrm{L0}} = f(\lambda) \label{eq:flamb} \\ 
        y_f(\lambda)&=& \frac{\lambda_0 f_if_\mathrm{L0}(\theta_i - \lambda/\Lambda) + \lambda_0 y_i
        f_\mathrm{L0}}{\lambda f_i + \lambda_0f_\mathrm{L0}}\\
        x_f(\lambda) & = & 0.
\end{eqnarray}
\end{subequations}
For the chosen geometry [Fig.~\ref{fig:off-axis zone plate schematic}(a)], the focal locations are confined to the $y$-$z$ plane (i.e., $x_f = 0$). However, the optical configuration is generally three-dimensional and can be rotated about the $z$ axis to move the focal locations into any other transverse plane. 

The longitudinal focal locations $z_f$ are determined by the combined focusing of the preliminary achromatic lens and diffractive lens. The focal lengths of these optics provide two parameters that can be tuned to set the overall f-number and longitudinal chromaticism. With $f_i$ and $f_\mathrm{L0}$ chosen, the transverse focal locations $y_f$ can be adjusted through the grating period $\Lambda$, incidence angle $\theta_i$, or initial transverse displacement $y_i$. The angles of and distances to a focus from the center point of the exit plane ($x=y=z^+=0$) are then 
\begin{subequations}
\label{eq:angledistance}
\begin{eqnarray}
 \theta_f(\lambda) &=& \theta_i - \frac{\lambda}{\Lambda} - \frac{\lambda}{\lambda_0}\frac{y_i}{f_\mathrm{L0}} \label{eq:labframeangle}
        \\
       d_{f}(\lambda) &=& \sqrt{y_f^2 + z_f^2},
\end{eqnarray}
\end{subequations}
where $\theta_f = \arctan{[(y_f - y_i)/z_f]} \approx (y_f - y_i)/z_f$ has been used in accordance with the paraxial approximation. 

\begin{figure}[t]
    \centering
    \includegraphics[width=1\linewidth]{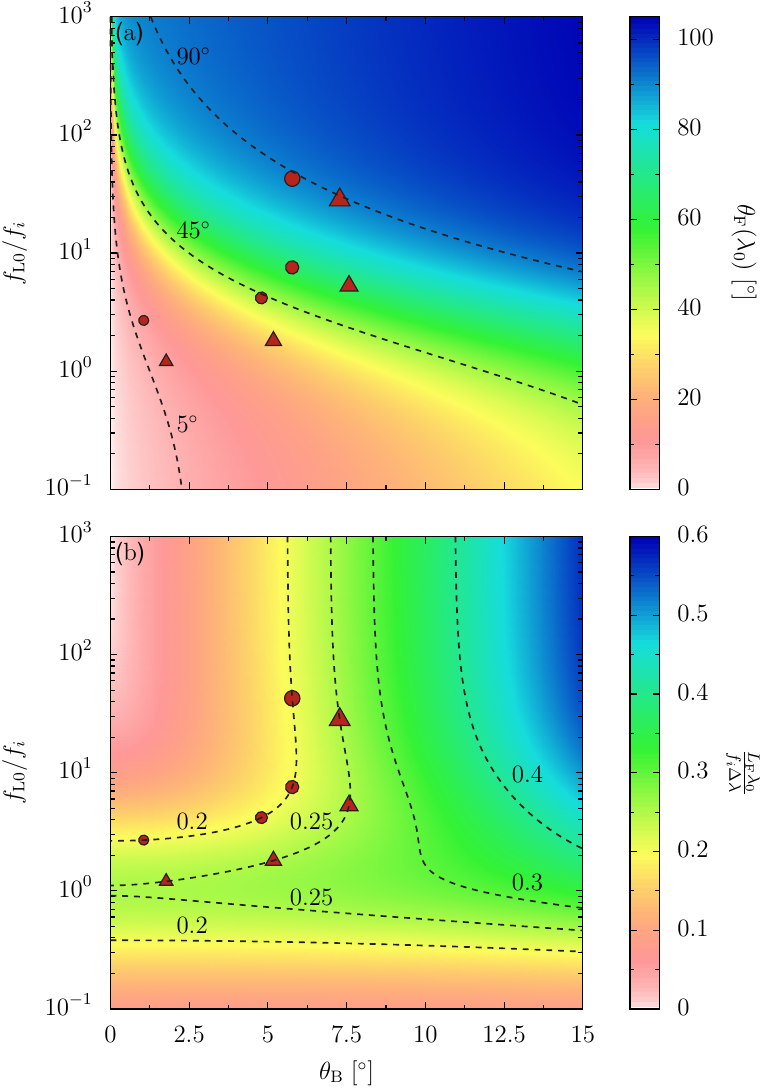}
    \caption{Design space for an arbitrary-directionality flying focus.
    (a) The angle 
    at the central wavelength with respect to its propagation direction
    and 
    (b) normalized focal range as a function of Bragg angle $\theta_\mathrm{B}$ and focal length ratio $f_\mathrm{L0}/f_i$.
    The contours show the analytical results [Eqs.~\eqref{eq:flying focus direction} and ~\eqref{eq:longitudinal displacement}] and the markers indicate the parameters used in the simulations.}
    \label{fig:analytic calculation of flying focus direction and distance}
\end{figure}

Equations~\eqref{eq:off-axis focal spot location and propagation direction} provide the focal locations in the natural coordinate system of the optical configuration. For analyzing the properties of the flying focus, it is convenient to instead use a coordinate system $(Y,Z)$ with an origin at the focal location of the central wavelength $\lambda_0$ and one axis ($Z$) aligned to the propagation axis of the central wavelength  [see Fig.~\ref{fig:off-axis zone plate schematic}(a)]. Within this coordinate system, the longitudinal and transverse displacements of the foci are given by
\begin{subequations}
\label{eq:longitudinal displacement}
    \begin{eqnarray}
        Z_\mathrm{F}(\lambda) &= &d_f\cos{(\theta_f - \theta_{f0})} - d_{f0}\\
         Y_\mathrm{F}(\lambda) &=& d_f\sin{(\theta_f - \theta_{f0})},
    \end{eqnarray}
\end{subequations}
where $\theta_{f0} \equiv \theta_f(\lambda_0)$ and $d_{f0} \equiv d_f(\lambda_0)$. For an incident pulse with a continuous spectrum, the displacements trace out a continuous curve or ``focal range'' of length
\begin{equation}
    \label{eq:flying focus range}
    L_\mathrm{F} = \int \sqrt{\left(\frac{dY_\mathrm{F}}{d\lambda}\right)^2 + \left(\frac{dZ_\mathrm{F}}{d\lambda}\right)^2}d\lambda, 
\end{equation}
where the integral is over all wavelengths composing the spectrum.

The motion of the focus through the focal range depends on the displacements $Z_\mathrm{F}(\lambda)$ and $Y_\mathrm{F}(\lambda)$ and the wavelength-dependent focal time $\tau(\lambda)$. The focal time $\tau(\lambda)$ consists of two contributions: the time it takes each wavelength to travel from $z = 0^+$ to its focal point, $d_f(\lambda)/c$, and the relative time of each wavelength within the pulse, $\partial_{\omega}\Phi$, such that
\begin{eqnarray}\label{eq:focaltime}
\tau(\lambda) = \frac{1}{c}d_f(\lambda) - \frac{\lambda^2}{2\pi c}\partial_{\lambda}\Phi,    
\end{eqnarray}
where $\partial_\omega\Phi = (\partial_\omega\lambda)\partial_\lambda\Phi$ was used to express the group delay in terms of wavelength. The focal velocities are then:
\begin{subequations}
\label{eq:flying focus velocity general pulse}
\begin{eqnarray}
    v_Z(\lambda) &=& \frac{dZ_\mathrm{F}}{d\lambda}\left(\frac{d\tau}{d\lambda}\right)^{-1}\\
    v_Y(\lambda) &=& \frac{dY_\mathrm{F}}{d\lambda}\left(\frac{d\tau}{d\lambda}\right)^{-1}.
\end{eqnarray}
\end{subequations}
Thus, the foci are dispersed at an angle $\theta_\mathrm{F} = \arctan(v_Y/v_Z)$ or
\begin{equation}
    \label{eq:flying focus direction}
    \theta_\mathrm{F}(\lambda) = \arctan\left[\frac{dY_\mathrm{F}}{d\lambda}\left(\frac{dZ_\mathrm{F}}{d\lambda}\right)^{-1}\right]
\end{equation}
with respect to the propagation direction of the central wavelength. Note that the angle is independent of $\tau$ and only depends on the lens and grating parameters.

Figure~\ref{fig:analytic calculation of flying focus direction and distance} shows how the angle and focal range of the flying focus depend on the design of the diffractive optic for a pulse with $y_i = 0$. To maximize the diffraction efficiency, the incidence angle of the pulse is fixed to the Bragg angle of the grating:  $\theta_i = \theta_\mathrm{B}$.
A purely longitudinal flying focus $\theta_\mathrm{F} = 0^\circ$ can be achieved when $\theta_\mathrm{B} = 0^\circ$, which is equivalent to only using a diffractive lens, as illustrated by Figs.~\ref{fig:off-axis zone plate schematic}(b) and (c).
A purely transverse flying focus $\theta_\mathrm{F} = 90^\circ$ can be achieved for large $f_\mathrm{L0}/f_i$, which is equivalent to only using a straight diffraction grating, as illustrated by Figs.~\ref{fig:off-axis zone plate schematic}(d) and (e). 
Angles between $0^\circ$ and $90^\circ$ can be achieved by choosing an appropriate combination of $\theta_\mathrm{B}$ and $f_\mathrm{L0}/f_i$, which corresponds to a diffractive lens and grating combination exemplified in Figs.~\ref{fig:off-axis zone plate schematic}(f) and (g). 
The focal spot aberrations visible in these figures arise from off-axis focusing and angular dispersion introduced by the diffraction grating~\cite{zhao2023investigation,zhao2024spatiotemporal}, but remain small in the paraxial regime considered here.
For a linearly chirped pulse with a stretched pulse duration $\tau_0$ and a bandwidth $\Delta\lambda/\lambda_0 \ll 1$, the velocity and direction of the flying focus are approximately constant in time. 
If the incidence angle is small such that $\theta_i^2 = \theta_\mathrm{B}^2\ll 1$, then $d_f(\lambda)\approx z_f(\lambda)$, and the focal range, velocities, and angle simplify to: 
\begin{subequations}
\label{eq:linearized flying focuse properties}
\begin{eqnarray}
    L_\mathrm{F} &\approx& \frac{\Delta\lambda}{\lambda_0} \left[4\theta_\mathrm{B}^2 + \frac{f^2(\lambda_0)}{f^2_\mathrm{L0}}\right]^{1/2}f(\lambda_0)\\
   v_Z &\approx&  c\left[1 \pm c\tau_0 \frac{\lambda_0}{\Delta\lambda}\frac{f_\mathrm{L0}}{f^2(\lambda_0)}\right]^{-1}\\
    v_Y &\approx& 2\theta_\mathrm{B}\left(1 + \frac{f_\mathrm{L0}}{f_i}\right)v_Z \\ 
    \theta_\mathrm{F} &\approx&\mathrm{arctan2}\,(v_Y, v_Z), \label{eq:flyingfocusAngleSimple}
\end{eqnarray}
\end{subequations}
where $\Phi = \pm (\pi c \tau_0/\Delta \lambda) (\lambda_0/\lambda - 1)^2$, the $\pm$ signs correspond to a positive and negative chirp, respectively, and $y_i = 0$ has been assumed.
In the limit of a purely longitudinal flying focus (i.e., $\theta_\mathrm{B} = 0$), the expressions for $L_\mathrm{F}$ and $v_Z$ agree with those derived in Refs.~\cite{froula2018spatiotemporal,sainte2017controlling,palastro2018ionization,li2024spatiotemporal}. 
Equation~\eqref{eq:linearized flying focuse properties} demonstrates that the focal length ratio $f_\mathrm{L0}/f_i$, grating period, and stretched pulse duration $\tau_0$, can be chosen to create a flying focus that travels at any speed and at any angle between $0^\circ$ and $180^\circ$.
A more general spectral phase could produce a focal point with a non-constant speed at angle $\theta_\mathrm{F}$.

For low-power applications, where laser damage is not a concern, a two-dimensional flying focus can be realized using a standard, solid-state diffractive lens and grating or a spatial light modulator. 
For high-power applications, the required optic could be imprinted holographically in a gas or plasma~\cite{michine2020ultra}. 
As shown in Ref.~\cite{edwards2022holographic}, imprinting the interference pattern of two co-linear laser beams with different focal lengths in a recording medium creates a holographic zone plate capable of
focusing a much higher power probe pulse. 
If the two imprint beams have the same wavelength $\lambda_I$ and focal points at $z = f_1$ and $z = f_2$, 
then the focal length of the holographic zone plate 
is
\begin{equation}\label{eq:hologramfz}
f_\mathrm{L}(\lambda) = \frac{\lambda_If_1f_2}{\lambda(f_1 - f_2)}.
\end{equation}
A holographic grating can be similarly created by crossing the two imprint beams at an angle in the recording medium.
If the two imprint beams cross at a full angle $2\alpha\ll 1$, then the grating period is $\Lambda = \lambda_I/2\sin\alpha \approx \lambda_I/2\alpha$. %
Crossing two imprint beams with different focal lengths at an angle in the recording medium creates a combined holographic zone plate and grating capable of applying a phase $\phi_\mathrm{L} + \phi_\mathrm{G}$ [Eqs.~\eqref{eq:quadratic phase} and \eqref{eq:phase diffraction grating}] to a probe beam.
The grating period and focal length can be controlled by changing the crossing angle and focal lengths of the imprint beams.

A combined holographic zone plate and grating can be created using a small volume of gas and integrated downstream of conventional focusing systems. 
Due to its high damage threshold, the optic can be placed close to the initial focal location. 
Reflection gratings and chirped mirrors that are common in high-power experiments can also be used to realize a flying focus as long as they apply the same phase $\phi_\mathrm{L} + \phi_\mathrm{G}$. 
However, unlike holographic optics, reflection gratings and chirped mirrors provide little to no tunability over the properties of the flying focus.

\begin{table}[t]
\caption{Simulation Parameters}
\label{tbl:params}
\begin{ruledtabular}
\begin{tabular}{l c c}
\noalign{\smallskip}
{\bf Paraxial}\footnote{The parameters are indicated in Fig.~\ref{fig:analytic calculation of flying focus direction and distance} by circles and triangles.}&\\
\noalign{\smallskip}
\hline
{Probe wavelengths $\lambda$}& $700$--$900\  \mathrm{nm}$\\
{Probe diameter}\footnote{The diameter is defined as the $1/e^2$ radius of intensity at the center of the medium.}& $300\ \mathrm{\upmu m}$ \\
{Probe incidence angle $\theta_i$}& $\theta_i = \alpha$\\
{Probe initial focal location $f_i$}& $12.0$ mm \\
{Imprint beam wavelength $\lambda_I$}& $800\ \mathrm{nm}$\\
{Imprint beam diameter}& $400\ \mathrm{\upmu m}$\\
{Imprint beam 1 incidence angle $\alpha$}& $1^\circ$--$7.5^\circ$\\
{Imprint beam 2 incidence angle $-\alpha$} & $-$\\
{Imprint beam 1 focal location $f_1$}& $154.7\ $mm\\
{Imprint beam 2 focal location $f_2$} 
& $13.4$--$116.7$ mm \\
{Index modulation amplitude $\delta n_\mathrm{A}$}& $10^{-4}$--$ 10^{-2}$\\
{Medium thickness $D$}& $ \lambda_0/\delta n_\mathrm{A}$ \\
{Parameter $\rho = \lambda_0^2/(\Lambda^2 n_0\delta n_\mathrm{A})$}& 1.8\\
\hline
\noalign{\smallskip}
{\bf PIC}&\\
\noalign{\smallskip}
\hline
{Probe central wavelength $\lambda_0$}& $800\  \mathrm{nm}$\\
{Probe bandwidth}& $161.6\  \mathrm{nm}$\\
{Probe duration $\tau_0$}\footnote{The probe is negatively chirped from $1.1\omega_0$ to $0.9\omega_0$ in $930\ \mathrm{fs}$.}& $930\ \mathrm{fs}$\\
{Probe diameter}& $150\ \mathrm{\upmu m}$\\
{Probe incidence angle $\theta_i$} & $10^\circ$\\
{Probe intensity $I$}& $5.4\times 10^{15}\ \mathrm{W/cm^2}$\\
{Probe initial focal location $f_i$}& $\infty$\\
{Imprint beam 1 incidence angle $\alpha$}& $5^\circ$\\
{Imprint beam 2 incidence angle} & $-5^\circ$\\
{Imprint beam 1 focal location $f_1$}& $\infty $\\
{Imprint beam 2 focal location $f_2$}& $750\ \mathrm{\upmu m}$\\
{Plasma thickness $D$}& $16\ \mathrm{\upmu m}$\\
{Plasma average density $N_0$}& $0.06$\\
{Plasma density modulation $N_1$}& $0.05$\\
{Ion mass $m_i/m_e$} & 1836\footnote{All ions are mobile in the simulation.}
\end{tabular}
\end{ruledtabular}
\end{table}

\begin{figure}[t]
    \centering
    \includegraphics[width=1\linewidth]{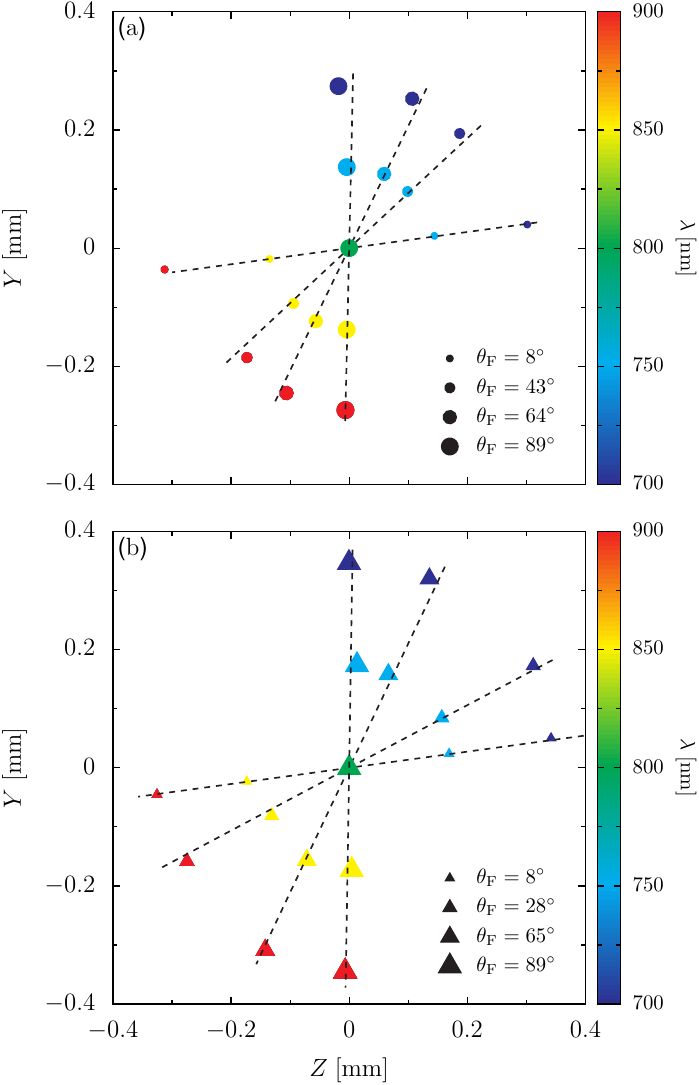}
    \caption{Focal location relative to the central wavelength $\lambda_0 = 800\ \mathrm{nm}$ for normalized focal ranges $L_\mathrm{F}\lambda_0/f_i\Delta \lambda$ equal to (a) 0.2 and (b) 0.25.
    The theory (dashed lines) agrees with the simulations (circles and triangles).
    The circles and triangles correspond to the same markers in Fig.~\ref{fig:analytic calculation of flying focus direction and distance}, which indicate the focal length ratios $f_\mathrm{L0}/f_i$ and Bragg angles $\theta_\mathrm{B}$ used in the simulations.}
    \label{fig:paraxial comparison}
\end{figure}

\section{Simulations}
We have demonstrated full control over the flying-focus direction using two complementary simulations: a numerical solver for the frequency-domain paraxial wave equation [Eq.~\eqref{eq:paraxial wave equation}] and time-domain particle-in-cell (PIC) calculations with the code EPOCH~\cite{Arber2015contemporary}. Both simulations model the propagation of a probe pulse from the near field, where it travels through a holographic lens and grating, to the far field, where it forms a flying focus. The paraxial solver also models the formation of the hologram by two imprint beams and allows for rapid calculations, but does not capture nonlinear feedback between the probe pulse and the recording medium. The PIC simulations do not model the hologram formation, but do capture the nonlinear feedback. The simulation parameters can be found in Table~\ref{tbl:params}. 
In the paraxial solver, the formation of the holographic optic and the propagation of the probe are performed in three steps:
(1) propagation of the imprint beams through a uniform medium with refractive index $n_0$;
(2) calculation of the refractive index modulation using the local interference pattern of the imprint beams: $\delta n(\mathbf{x}) = n(\mathbf{x}) - n_0 \propto \mathbf{E}_1 \cdot \mathbf{E}_2$;
(3) propagation of the probe through the resulting nonuniform medium with refractive index $n(\mathbf{x})$.
 For all cases considered, the incidence angle of the probe was equal to the Bragg angle $(\theta_i = \theta_\mathrm{B})$. 
The focal range $(L_\mathrm{F})$ and angle $(\theta_\mathrm{F})$ of the flying focus were varied by changing the focal length of the second imprint beam $f_2$ and the crossing angle of the imprint beams $2\alpha$, which set $f_\mathrm{L}$ [Eq.~\eqref{eq:hologramfz}] and the Bragg angle: $\theta_\mathrm{B} \approx \lambda_0/(2\Lambda)$ with $\Lambda = \lambda_I/(2\sin\alpha)$, respectively. The amplitude of the index modulation $\delta n_\mathrm{A} = \mathrm{max}(\delta n)$ was increased with the Bragg angle to mitigate higher-order diffraction. More specifically, the parameter $\rho = \lambda_0^2/(\Lambda^2 n_0\delta n_\mathrm{A}) = 1.8 > 1$ was held fixed~\cite{moharam1978criterion}, resulting in a single bright, first-order focus. The values of $\delta n_\mathrm{A}$ ranged from $10^{-4}$ and $10^{-2}$, well within the range achievable in gas and plasma~\cite{edwards2022holographic,Edwards2022plasma,michel2024photochemically}. The thickness of the hologram $D$ was chosen to satisfy $(D / \lambda_0)\delta n_\mathrm{A}  = 1$, which yields maximum diffraction efficiency at the central wavelength~\cite{edwards2022holographic}.

Figure~\ref{fig:paraxial comparison} compares the results of the paraxial simulations (circles) to the theory  (dashed lines) for probe wavelengths ranging from $700$ to $900\ \mathrm{nm}$. The simulated  longitudinal and transverse focal point displacements match the analytic prediction given by Eq.~\eqref{eq:longitudinal displacement} across a broad range of focal length ratios $f_\mathrm{L0}/f_i$ and Bragg angles $\theta_\mathrm{B}$. 
The focal location in the simulations was determined by finding the maximum intensity $\propto|\tilde{E}|^2$ in the first-order diffraction. For these parameters, the focal ranges (a) $L_\mathrm{F} \approx 600\ \mathrm{\upmu m}$ and (b) $L_\mathrm{F} \approx 700\ \mathrm{\upmu m} $ are $1.5$--$5$ and $1.75$--$6$ times the Rayleigh range of the focused probe, respectively. Since $L_\mathrm{F}$ scales with $f_i$ at fixed $f_\mathrm{L0}/f_i$, the focal range can be substantially increased by moving the holographic optic further upstream from the initial focal location. The velocity of the focus through the focal range is determined by the spectral phase, which is left unspecified here for generality. In the case of a linear chirp, for instance, the sign of the spectral phase determines whether the focus is moving forward or backward. 

\begin{figure}[t]
    \centering
    \includegraphics[width=1\linewidth]{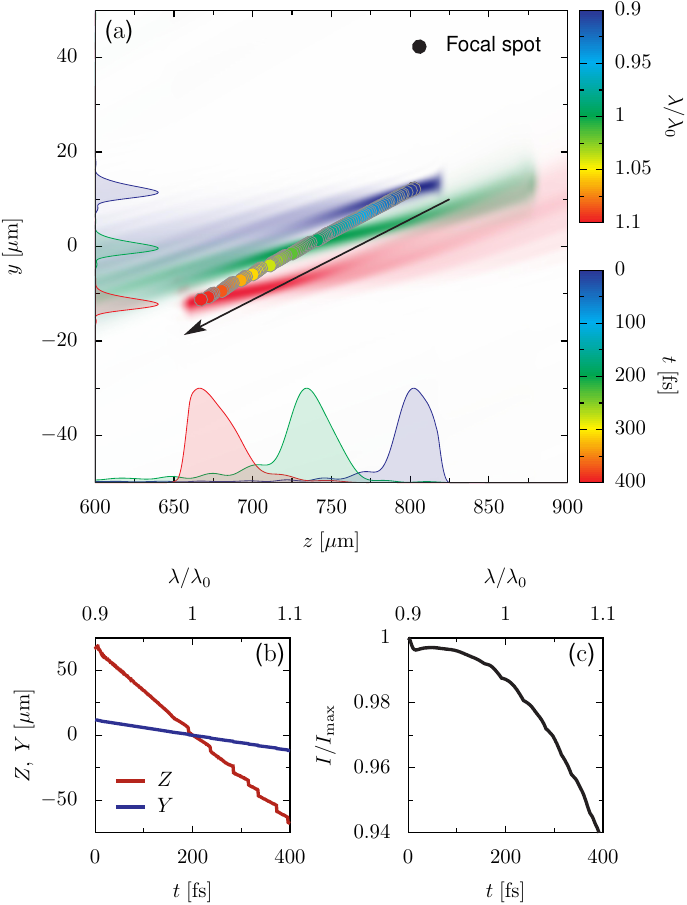}
    \caption{PIC simulation of a two-dimensional flying focus produced by focusing and diffracting a chirped laser pulse with a combined plasma zone plate and grating.
    (a) The pulse intensity at three different times, corresponding to the moments at which light at different wavelengths reach focus, showing that the focal spot moves at an angle with respect to the propagation direction of the central wavelength. The dots show the focal location as a function of time.
    (b) The longitudinal and transverse displacement with respect to the focal location of the central wavelength as a function of time and wavelength.
    (c) The focused intensity as a function of time and wavelength.}
    \label{fig:pic simulation}
\end{figure}

The paraxial simulations verify the theoretical predictions provided that nonlinear feedback between the probe pulse and the holographic optic is negligible. Thus, they apply only to probe pulses whose intensities are too low to modify the recording medium. Plasma can withstand high intensities, making it a promising media for holographic optics. Two-dimensional PIC simulations, which solve the full set of Maxwell's equations and capture nonlinear feedback, were conducted to demonstrate the formation of a high-intensity, arbitrary-directionality flying focus using a plasma zone plate and grating. 

For the PIC simulations, the parameters of the probe pulse, plasma zone plate, and grating were designed to produce a constant-velocity flying focus with $L_\mathrm{F} = 150\ \mathrm{\upmu m}$, $v_Z = -1.16c$, and $v_Y = - 0.20c$. A linearly chirped probe pulse with an intensity $I \approx 5.4\times 10^{15}\ \mathrm{W/cm^{2}}$ (normalized vector potential  $a_0 = \mathrm{max}(E)e/m_e\omega_0c = 0.05$) was obliquely incident on a combined plasma zone plate and grating at an angle $\theta_i = 10^\circ$. 
The temporal profile of the chirped probe consisted of a 24 fs linear rise, a 930 fs plateau, and a 24 fs linear fall. 
The simulations had a resolution of 32 cells/$\lambda_0$ and used 1 macroparticle per cell. A series of simulations conducted with higher resolution and more macroparticles per cell confirmed that these values were sufficient for convergence.  The simulation domain was 0.6 mm $\times$ 1 mm in $y$ $\times$ $z$.

The preformed plasma zone plate and grating were implemented as an initial, neutral density profile: 
\begin{equation}
N(y,z) = N_0 + N_1\cos[\Delta \phi(y,z)], 
\end{equation}
where $N_0$ and $N_1$ are the average density and density modulation normalized by the critical density $n_c = \epsilon_0m_e\omega_0^2/e^2$ and 
\begin{equation}
\Delta\phi(y,z) = \frac{k_Iy^2}{2(z - f_2)} - 2k_Iy\sin\alpha
\end{equation}
is the relative phase between two imprint beams. The corresponding refractive index is $n(y,z) = \sqrt{1 - N(y,z)}$. The parameters of the density profile (see Table~\ref{tbl:params}) were motivated by ponderomotive plasma gratings, where the intensity modulation due to the interference of two imprint beams produces a ponderomotive force that pushes plasma into regions of low intensity~\cite{sheng2003plasma, Lehmann2016transient, edwards2022holographic}.

Figure~\ref{fig:pic simulation}(a) shows the probe pulse intensity after passing through the plasma hologram at three times, corresponding to the focal times of three wavelengths (or time slices) within the pulse (see colorbar). 
Each wavelength travels in a different direction and focuses at a distinct location and time [Eqs.~\eqref{eq:off-axis focal spot location and propagation direction} and \eqref{eq:focaltime}]. 
Figure~\ref{fig:pic simulation}(b) demonstrates that the focal point has a near-constant velocity in both the longitudinal and transverse direction: $v_Z\approx -1.1c$ and $v_Y\approx -0.195c$ and moves backward at an angle $\theta_\mathrm{F}\approx 10^\circ$ with respect to the pulse propagation direction, consistent with Eq.~\eqref{eq:flyingfocusAngleSimple} ($\theta_\mathrm{F}\approx 10^\circ$).
The simulated velocities slightly differ from the predictions of Eq.~\eqref{eq:linearized flying focuse properties}.
This is due to the small-angle and small-bandwidth approximations used to derive Eq.~\eqref{eq:linearized flying focuse properties}.
Despite these differences, Eq.~\eqref{eq:linearized flying focuse properties} provides a quick and reasonably accurate estimate of the focal speed and direction.
The peak intensity of the moving focus varies by about 6\% across the entire focal range, reaching a maximum of $6.4\times 10^{16}\ \mathrm{W/cm^2}$ ($a_0\approx 0.17$) [Fig.~\ref{fig:pic simulation}(c)].
These results indicate that the holographic plasma optic has a damage threshold at least three orders of magnitude higher than that of conventional solid-state optics ($10^{12}\ \mathrm{W/cm^{2}}$). 
This capability positions holographic plasma optics as an alternative to large solid-state optics for the manipulation of high-power laser pulses.

\section{Conclusion}
In conclusion, we have introduced an arbitrary directionality flying focus that conceptually advances the flying focus technique to two dimensions. 
This design allows the focal trajectory to be steered at an angle with respect to the pulse propagation direction while preserving the extended interaction length and arbitrary focal velocity that make the flying focus technique attractive. 
The method decouples the longitudinal and transverse motion of the focal point by using a diffractive lens and grating to focus and diffract a broadband, chirped laser pulse.
The optical configuration can be realized using conventional, adaptive, metasurface, or photorefractive optics, providing a range of options depending on the intensity that is needed. 
The combined optic affords greater control over the focal-point motion beyond what is achievable with existing methods, without adding substantial complexity to the optical setup.
The ability to independently control the direction and velocity of the intensity peak allows the interaction geometry to be tailored without sacrificing the extended interaction region. Moving the focal spot laterally with respect to the pulse propagation direction reduces the risk of optical damage to upstream and downstream optical components, improves shielding and diagnostic access, and provides greater flexibility in experimental design. These practical considerations are particularly important in high-power laser–plasma experiments, where geometric constraints often limit implementation of otherwise promising concepts.

\bigskip

\noindent {\bf Acknowledgments. }This work was partially supported by NSF Grant PHY-2308641 and NNSA Grant DE-NA0004130. The work of J.P.P. and L.M. was supported by the Department of Energy Office of Science under Award Numbers DE-SC0021057 and DE-SC0025497/SUB0000841 and the Department of Energy National Nuclear Security Administration under Award Number DE-NA0004144. The PIC code EPOCH~\cite{Arber2015contemporary} is funded by the UK EPSRC grants EP/G054950/1, EP/G056803/1, EP/G055165/1 and EP/ M022463/1. The computing for this project was performed on the (Stanford) Sherlock cluster. We would like to thank Stanford University and the Stanford Research Computing Center for providing computational resources and support that contributed to these research results. \\

\noindent {\bf Disclosures.} The authors declare no conflicts of interest.\\

\noindent {\bf Data availability. }Data underlying the results presented in this paper are not publicly available at this time but may be obtained from the authors upon reasonable request.

\end{document}